\documentclass[a4paper]{spie}  
\usepackage{amsmath,amsfonts,amssymb}

\usepackage[utf8]{inputenc}   
\usepackage[english]{babel}   
\usepackage[usenames,dvipsnames]{xcolor} 
\usepackage[colorlinks=true, allcolors=blue, pdftex]{hyperref}
\usepackage{aas_macros}
\usepackage[free-standing-units,binary-units]{siunitx}
\usepackage{paralist}
\usepackage{graphicx}
\usepackage{multicol}
\usepackage[frozencache,cachedir=.]{minted}

\title{Easy remote observations using web interfaces: \\ controlling
  an Italian telescope from Japan, and more }

  \author[a,*]{Davide~Ricci}
  \author[b]{Lorenzo~Cabona}
  \author[a]{Bernardo~Salasnich}
  \author[c]{Luciano~Nicastro}
  \author[d]{Luca~Fini}
  \author[e]{Andrea~Damonte}
  \author[f,g,b]{Silvano~Tosi}
  \author[h]{Takashi~Shibata}
  
  \affil[a]{INAF-Osservatorio Astronomico di Padova, Vicolo dell'Osservatorio 5, 35122 Padova, (Italy).}
  \affil[b]{INAF - Osservatorio Astronomico di Brera, Via E. Bianchi 46, 23807, Merate, (Italy).}
  \affil[c]{INAF - Osservatorio di Astrofisica e Scienza dello Spazio, via P. Gobetti 93/3, I-40129 Bologna, (Italy).}
  \affil[d]{INAF - Osservatorio Astrofisico di Arcetri, Largo Enrico Fermi, 5, I-50125 Firenze, (Italy).}
  \affil[e]{INAF - Osservatorio Astronomico di Palermo "Giuseppe Salvatore Vaiana", Piazza del Parlamento 1, 90134 Palermo (Italy).}
  \affil[f]{Università degli Studi di Genova, DIFI Dipartimento di Fisica, Via Dodecaneso 33, 16146, Genova, (Italy).}
  \affil[g]{INFN-Sezione di Genova, Via Dodecaneso 33, 16146 Genova, (Italy).}  
  \affil[h]{National Astronomical Observatory of Japan, 2 Chome-21-1 Osawa, Mitaka, Tokyo 181-8588 (Japan).}
  
  \authorinfo{$^*$Contact information: davide.ricci@inaf.it}

\begin{document}
\maketitle

\renewcommand{\thefootnote}{\alph{footnote}}

\begin{abstract}
  Remote observations are often limited by user interfaces, which seem
  frozen to another computer era: low performances, outdated
  programming languages, command-line scripting, high
  version-dependent software. Instead, web-based tools are standard:
  using nothing more than a browser, astronomers can interact with a
  generic observatory in a native cross-platform, remote-born way. We
  used this approach while advancing in the remotization of the
  1m-class OARPAF telescope, located in Northern Italy. The web-based
  control software provides easy and integrated management of its
  components. This solution can be exported not only to similar
  hardware/software facilities, but also to large instruments such as
  SHARK-NIR at LBT; not only for single operations, but also for
  procedure scripting.  In this contribution we describe our best
  practices and present two recent, orthogonal use cases: an in-place
  professional use for exoplanetary transit follow-ups, and the first
  remote control of the telescope from a Japanese high school,
  allowing students to independently observe, in their daytime,
  globular clusters.
\end{abstract}


\section{Introduction}
\label{sec:intro}

Instrument Control Software (INS) is one of the key components of
astronomical instrumentation. It manages and coordinates all
instrument components, such as: motors, calibration lamps, eventual
Adaptive Optics modules, image generation and reconstruction, and
sometimes the telescope interface and the image archiving process.

Large international organizations, such as ESO or ALMA, developed
speciﬁc frameworks which are used for the development of the INS of
their instrumentation, while other observatories, such as LBTO, use
more flexible approaches.
The latter allows us to experiment new solutions, both tailored on the
characteristics of a speciﬁc instrument or inherited from common
components.

It is the case of LBTI and the recently installed SHARK-NIR
instruments at LBT. They share the {\tt TwiceAsNice} framework
developed by MPIA for motors, but use separate, custom software
developed within INAF for the control of the remaining
components\cite{2022SPIE12189E..20R}.

Concerning control interfaces, LBTO is moving forward to web
applications for its telescope pointing system, and suggested the web
way also to control SHARK-NIR. Recent updates are being presented in
this conference\cite{lorenzetto-shark}, and are based on an approach
developed and tested on the OARPAF (Osservatorio Astronomico Regionale
Parco Antola Comune di Fascia) $1\meter$ class telescope in Northern
Italy\cite{2022SPIE12186E..0PR}.

Here we present the next step of this journey: a software
generalization that we hope will encourage its adoption in other
similar-class facilities.
In the first part we show how to set up from scratch the control
software of a dummy observatory by editing configuration files; it can
then be operated locally using Python scripts and remotely using REST APIs.
In the second part we show how we implemented the control software at
OARPAF, building on top of this a web-based control interface.
Then, we show a use case of remote control.

\section{Configuring an observatory control software from scratch}

\begin{figure} [t]
  \centering
  \includegraphics[width=\textwidth]{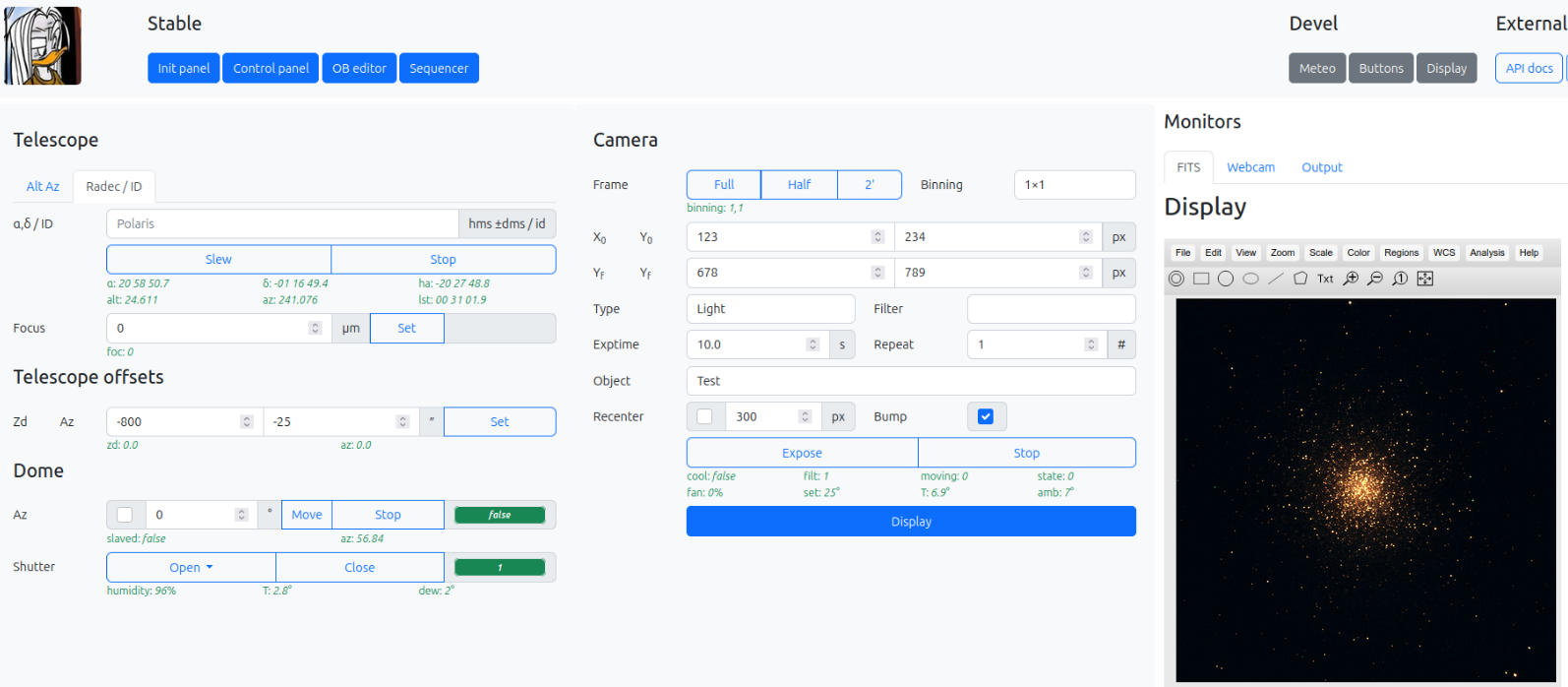}
  \label{fig:panel} 
  \caption{ The OARPAF Control panel.  }
\end{figure}

The control software concept presented in this contribution was
initially tailored on the OARPAF available elements. In particular,
limiting to the main devices:
\begin{inparaenum}
\item the telescope control is based on \texttt{telnet} commands
  following the proprietary \texttt{OpenTSI} protocol. A proprietary
  \texttt{ASCOM} driver for the telescope was also purchased from the
  Astelco telescope company;
\item the dome control is based on the proprietary \texttt{OmegaLab}
  software, which is also exposed as \texttt{ASCOM} device;
\item the imaging control system (CCD and filter wheel), and the
  separate monitoring IP camera, are based on HTTP APIs provided by
  the respective vendors.
\item \texttt{SonOff} switches allow the CCD camera power to be
  controlled via HTTP APIs.
\end{inparaenum}
On the basis of this, we initially used the \emph{ASCOM Remote}
component, which is essentially a web server exposing \texttt{ASCOM}
methods as REST APIs in a parallel standard called \texttt{Alpaca},
and we interfaced telescope and dome to it.  We then built: a
\texttt{python} layer to control all these individual devices; a
series of scripts to group common operations in a ``ESO-style''
Templates and Observation Blocks (OBs) pattern; finally we built our
own common REST APIs for all the devices.  On top of this, we
developed web panels to control the devices (see
Fig.~\ref{fig:panel}), and to edit and launch OBs.  Details of these
steps are given in our previous
contribution\cite{2022SPIE12186E..0PR}.

The first upgrade came from the idea to get rid of the buffer Windows
pc that run the \emph{ASCOM Remote} server; then we developed our own
\texttt{OpenTSI}-based \texttt{python} interface to the telescope,
instead of relying on the ASCOM driver.  With the same aim, we adopted
a custom \texttt{python} daemon to control the dome hardware, exposed
as a native \texttt{Alpaca} device, which was initially developed for
a same class dome in use at the OPC observatory.
Effort are in progress to support the OPC telescope, based on
\texttt{LX200} commands, and to develop a common software layer for
both observatories.
Our goal is to try to involve observers who can share compatible or
totally different components within this project. Among the first we
could typically list the dome electronics and the related software. On
the other hand, scientific cameras have their own specific protocols
and interfaces.  Involving as many contributors as possible would
allow us to build increasingly generic interfaces and consequently
expand the user community.

Following this pattern, the code provides an easy customization via
\texttt{INI} configuration files. In this example, we assume to
configure an observatory composed by
\begin{inparaenum}
\item an \texttt{ASCOM} dome;
\item an \texttt{ASCOM} telescope,
\item a \texttt{SonOff} switch.
\end{inparaenum}
We first fill the \texttt{config/nodes.ini} configuration file, to
specify where the devices can be found.  In our example, two different
addresses are inserted. Then, in the \texttt{config/devices.ini}
configuration file, we associate each device to a node, specifying the
device type and class (for example an ``Alpaca Telescope''):

\begin{multicols}{2}
  
\begin{minted}{ini}
# config/nodes.ini

[MY_ASCOM_REMOTE_SERVER]
protocol = http
ip = 192.168.1.111
hostname = ascom.myobservatory.net
port = 533
endpoint = /api/v1

[MY_SONOFF_ADDRESS]
protocol = https
user = admin
password = admin
ip = 192.168.1.222
hostname = sonoff.myobservatory.net
port = 1234
\end{minted}
  
\begin{minted}{ini}
# config/devices.ini

[my_tel]            
module = alpaca 
class = Telescope
node = MY_ASCOM_REMOTE_SERVER   

[my_dome]            
module = alpaca  
class = Dome    
node = MY_ASCOM_REMOTE_SERVER

[my_light]
module = domotics
class = Switch
node = MY_SONOFF_ADDRESS
outlet = 3
\end{minted}

\end{multicols}

\noindent With this simple configuration, it is already possible to
control the devices using a \texttt{python} script, or directly from a
\texttt{python} console, using a series of \emph{pythonic}
getter/setter methods:
\begin{minted}{python}
import devices
devices.my_dome.azimuth # gets current value, i.e. 123
devices.my_dome.azimuth = 234 # moves the dome to the new value
\end{minted}
Scripts can also be organized into \texttt{Template} classes,
inheriting from the \texttt{BaseTemplate} class. This allow us to
invoke them from the \texttt{sequencer.py} script, passing parameters
in \texttt{JSON} Observation Blocks that may contain one or more
template instances.
In the following example, we provide a very dummy Template script to
change the dome azimuth, called \texttt{templates/change\_az.py}.  Of
course, scripts are expected to manage more complex, multi-device
operations.  Then, we show the corresponding Observation Block
parameter file (\texttt{ob/my\_azimuth\_update.json}), and the
sequencer call:

\begin{multicols}{2}
  
\begin{minted}{python}
# templates/change_az.py
import devices
from templates.basetemplate import BaseTemplate

class Template(BaseTemplate):
    def content(self, params):
        current_az = devices.my_dome.azimuth
        next_az = params["azimuth"] # 234
        devices.my_dome.azimuth = next_az
\end{minted}

\begin{minted}{json}
// ob/my_azimuth_update.json
[
  {"template" : "change_az",
   "params": { "azimuth": 234 }
  }
]
\end{minted}

\begin{minted}{bash}
# Running the Observation Block in the console
./sequencer.py ob/my_azimuth_update.json
\end{minted}
  
\end{multicols}

\noindent This series of examples show a basic use case of the
control software concept.  It allows us to operate locally at the device
level using a python console, or building and run custom scripts,
or building Template-like scripts and running them using a sequencer.
Of course, as each observatory has its own configuration, all
templates must be customized.

Another control layer allows us to configure REST APIs for remote control
at device or template level. It is sufficient to edit an additional
configuration file, \texttt{config/api.ini}, choosing the desired
endpoints and specifying the API resource with the corresponding
device.  This layer allows us not only to remotely control devices with
an HTTP call, but also adds additional information in a \texttt{JSON}
response, with respect to the \texttt{python} layer, such as grouped
values, a timestamp and a dictionary-parsed output.  Once completed,
endpoints can be accessed after starting the API server with the
command \texttt{./app.py 1111 localhost --noweb}. In the following
example we suggest some endpoints, we show the output of a getter
using the \texttt{curl} UNIX command, an example of a setter command,
and a specific ``bulk GET call'' that retrieves all endpoints of a
first level nesting in once:

\begin{multicols}{2}

\begin{minted}{ini}
# config/api.ini

[/dome/lamp]
resource = State
device = my_light
priority = 1

[/alias/to/the/switch]
resource = State
device = my_light

[/dome/position]
resource = Position
device = my_dome

[/telescope/coordinates]
resource = Coordinates
device = my_tel

\end{minted}

\begin{minted}{bash}
# Get azimuth
curl http://localhost:1111/api/dome/position

# Returned JSON example:
# {"timestamp" : "2024-03-26T16:48:13.766195"
#  "error" : [],
#  "raw" : False, # is parked?
#  "response" : {
#     "azimuth" : 123.00,
#     "parked" : "No"
#   },
# }

# Change azimuth
curl -X POST http://localhost:1111/api/dome/position \
     -d '234' -H 'Content-Type:application/json'

# Retrieve all GET endpoints relative to the dome 
curl http://localhost:1111/all/dome
\end{minted}

\end{multicols}

\noindent The next step foresees to generalize, via configuration file,
also a web snippet for each API resource, in order to provide an easy
set up of a default web page.

In the case of OARPAF, we built a custom web page. We use the ``bulk
GET endpoints'' to build \texttt{JSON} statuses that are periodically
sent to the web page using websockets. Then, we interrogate APIs for
the control.
FITS files obtained by the camera are currently visualized using
\texttt{JS9}, which sends data in binary format and renders them with
a web worker in a html canvas.

We are developing a \texttt{WebGL} custom solution, and we will test
sending the data as an \texttt{ArrayBuffer}, omitting the least
significant byte of each pixel to reduce the transfer size,
sacrificing some display counting accuracy for speed purposes.

In the next Section we show an example of remote control at OARPAF
using the custom web interface.

\section{Remote Control use case}
\label{sec:devs}

\begin{figure} [h]
  \centering
  \includegraphics[width=0.55\textwidth]{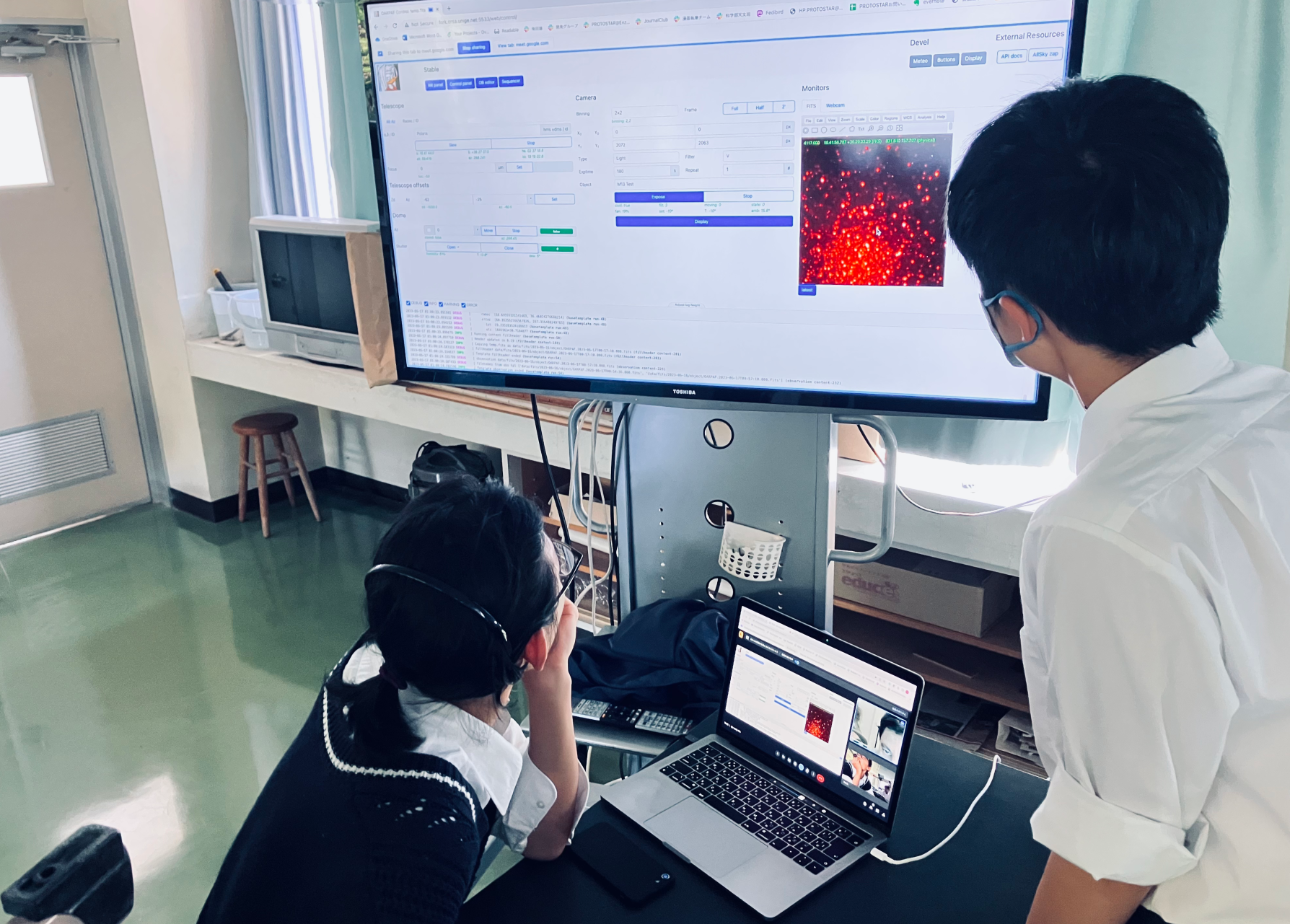}
  \includegraphics[width=0.44\textwidth]{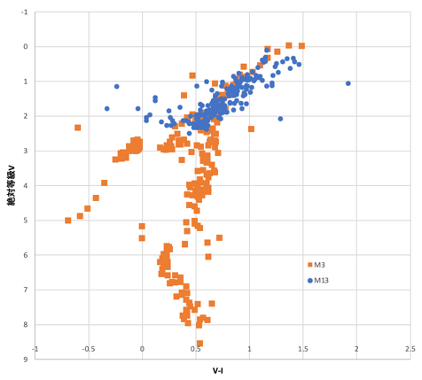}
  \label{fig:cmd} 
  \caption{Left: Remote OARPAF observations driven by K\=ory\=o High
    School students. Right: Color-Magnitude produced by the
    students; blue: M13 obtained at OARPAF; orange: literature data. }
\end{figure}

K\=ory\=o High School is an institute of the Hokuto city in the
Yamanashi prefecture, offering astronomy as side activity in the
framework of its ``Science Club''.
We organized a remote observation at OARPAF, conducted by several
students directly from their classroom.  Students preliminarly
discussed the science case, choosing the multi-filter study of the
Globular Cluster M13, and formalised it into an ``observation proposal''
complemented by a visibility chart and the scientific goals.

They then connected at \texttt{8:00 JST}, corresponding to
\texttt{23:00 UTC}, or \texttt{1:00 CET}, using a 4G mobile hotspot to
optimize internet connection speed.  Then, they remotely controlled
the telescope cover, the dome shutter, the telescope pointing. After
that they run a focus procedure, and finally took three $180\second$
frames in the $V$ filter and three $180\second$ frames in the $I$
filter, displaying each FITS file in the \texttt{JS9} instance of the
control web page, all with a limited lag due to the data file transfer
(around $30\second$ for each $33\unit{\mega\byte}$ of a full frame
image).
Operations ended at sunrise, i.e. around \texttt{12:00 JST = 03:00 UTC
  = 05:00 CET}.

They decided to carry out a very rough photometry estimation of
$\approx 200$ stars using the ring options of \texttt{ds9}, and
choosing as photometric standards isolated bright stars of the cluster
for which the Simbad catalog provided both $V$ and $I$
magnitude. Finally, they compared the results in a Color-Magnitude
Diagram (see Fig.~\ref{fig:cmd}) with respect to literature
data\footnote{\url{https://www.astroarts.co.jp/alacarte/messier/index-j.shtml}}
of another Globular Cluster, M3.

The students presented the results of the observing night at the
\emph{44th Yamanashi Prefecture High School Arts and Culture Festival
  Natural Science Division ``Student Natural Science Research
  Presentation Competition''}, and have been awarded with the second
place over 50 groups and a total of 200 high school participants.
This allowed us to validate our web-based approach for a
$\approx 10\,000 \,\km$ distance use case, confirming its usability
and paving the foundation for a further iteration of the
collaboration, and in general for activities in the third mission
sector.

\section{Conclusion}
\label{sec:conc}

In this work we presented the progresses in generalizing the control
software we are implementing for the OARPAF 1m-class observatory
towards other observatories.
The main effort consisted in providing a simple set of configuration
files to set up supported devices ad control them locally using
\texttt{python} (individually or using OBs based on Template scripts),
and remotely using REST APIs.
We developed a custom web-based client interface for OARPAF, based on
websockets for periodic status check and API calls for control. We aim
at providing in the future a common, default web control interface
customizable by editing a configuration file.
This remote control system has been successfully operated by Japanese
High School students in the framework of a science project, which
confirms it to be a valid approach not only for a professional use but
also for outreach activities.

\acknowledgments DR acknowledges Comitato Interministeriale per la
Programmazione Economica (\texttt{C93C23008400001}).

\bibliographystyle{spiebib} 
\bibliography{biblio} 

\end{document}